\documentstyle[preprint,aps,epsf,floats,eqsecnum]{revtex}

\def\bccs{b\rightarrow c\overline{c}s}
\def\bcud{b\rightarrow c\overline{u}d}
\def\bctaunu{b\rightarrow c\tau\overline{\nu}_\tau}
\def\bxctaunu{\bctaunu}

\def\CO{{\cal O}}
\def\OMIT#1{{}}
\begin{document}

\preprint{\vbox{\hbox{UTPT--96-07}
\hbox{CMU--HEP96--07}
\hbox{DOE--ER/40682--118}
\hbox{hep-ph/9605406}}}

\title{Higher Order QCD Corrections to $b\rightarrow c\overline{c}s$}

\author{Ming Lu$^a$, Michael Luke$^b$, Martin J. Savage$^{a}$
\footnote{Address after Sept. 96: Dep't of Physics, University of
Washington, Seattle WA 98195} and
Brian H. Smith$^b$}
\address{
$^a$Department of Physics, Carnegie Mellon University\\
Pittsburgh, Pennsylvania 15213 U.S.A\\
$^b$ Department of Physics, University of
Toronto\\ 60 St.~George Street, Toronto, Ontario, Canada M5S 1A7}

\date{May 1996}

\maketitle

\begin{abstract} We calculate the ${\cal O}(\alpha_s^2 \beta_0 )$ corrections
to the
decay
rate $b\rightarrow c\bar c s$.   For reasonable values of $m_c/m_b$ this
term is of the same order as both the one-loop and $\CO(\alpha_s^2 \log^2
m_W/m_b)$
corrections to the decay rate.   For $m_c/m_b=0.3$ the $\CO(\alpha_s^2\beta_0)$
corrections enhance the
rate by $\sim 18\%$ . We also discuss the
$\CO(\alpha_s^2\beta_0)$ corrections to $R_\tau$, the $B$ semileptonic
branching
fraction
and the charm multiplicity.

\end{abstract}


\section{Introduction}

The doubly-charmed decay mode of the $B$ meson, $B\rightarrow
X_{c\bar c s}$, has been the object of recent interest, since
this mode makes a significant contribution to the inclusive $B$
semileptonic branching fraction \cite{semitheory,CLEO}.
Recently, the one-loop
corrections to $\bccs$ were calculated \cite{mv,bbfg} (see also \cite{hp}) 
and found to be substantial, giving
a $\sim 22\%$ enhancement to the tree level rate (for $\hat m_c=0.30$).  This
is significantly larger
than the corresponding $\sim 5\%$ $\CO(\alpha_s)$ correction to $\bcud$ decay.
When combined with the additional radiative corrections, this brings the
theoretical prediction for the semileptonic branching fraction into agreement 
with the experimental observation, within
the theoretical uncertainties \cite{neubert96}.

Since the typical energy released in the decay,
$\Delta\equiv m_b-2 m_c$ (neglecting the $s$ quark mass), is much less than
$m_b$,
one might expect the relevant scale for the perturbative corrections to
$\bccs$ to be significantly less
than $m_b$.  Indeed, as stressed in Ref.\ \cite{FWDa}, the energy release in
this process is
so small that the assumption of local duality may not hold; it has
been argued in Ref.\ \cite{shifman94} that deviations from duality would not
show up at any
finite order in the operator product expansion.  However, even if the
assumption of
local duality does hold in this instance, this low scale would result in an
even greater
enhancement of this mode over the tree-level result.  This is a
higher order effect which requires a full two-loop calculation to address,
which we have not attempted.  However, in the approach of Brodsky, Lepage and
Mackenzie
(BLM) \cite{BLMa} useful
information may be obtained by simply calculating the $n_f$ dependent piece of
the order
$\alpha_s^2$ contribution to the decay.
This determines the contribution
of ${\cal O}(\alpha_s^2\beta_0 )$, where $\beta_0 = 11-{2\over 3}n_f$.  Since
$\beta_0$ is large, this term dominates the two loop result for many processes.
The BLM scale $\mu_{\rm BLM}$ for the one-loop correction is defined as the
scale
at which the $\CO(\alpha_s^2\beta_0)$ correction is absorbed in the one-loop
correction.
This approach has recently been used to estimate the
two-loop corrections to semileptonic top, bottom and charm decays
\cite{SmVol,LSW}.

In this paper we calculate the
$\CO(\alpha_s^2\beta_0)$ correction to the decay $\bccs$.
We will find that this term enhances the decay rate by almost as much as the
one loop term, and is of the same size as the $\CO(\alpha_s^2\log^2 m_W/m_b)$
correction.   However,
as we will discuss, the $\CO(\alpha_s^2\beta_0)$  term is not necessarily
expected
to dominate the
remaining uncomputed two-loop corrections.

In Section 2 we compute the ${\cal O}(\alpha_s^2\beta_0)$ corrections to the
mode
$\bctaunu$. This contribution arises from strong interaction corrections to the
$bc$
vertex, and the result can be related to a piece of the
$\bccs$ correction by taking $m_\tau= m_c$.  These corrections are
interesting in their own right as they give $\CO(\alpha_s^2\beta_0)$
corrections to
the ratio
\begin{equation}\label{rtau}
R_\tau\equiv{\Gamma(b\rightarrow X_c\tau\bar\nu_\tau)\over
\Gamma(b\rightarrow X_c e\bar\nu_e)}.
\end{equation}
We will find that the two-loop corrections to this ratio are under control.
In Section 3 we calculate the $\CO(\alpha_s^2\beta_0)$ corrections to the
$\bar c s$ vertex.   We give our conclusions in Section 4.

\section{${\cal O}(\alpha_s^2\  \beta_0)$ corrections to $\bctaunu$}
\medskip

The rates for $B\rightarrow X_c e\bar\nu_e$ and $B\rightarrow X_c
\tau\bar\nu_\tau$
may be written as power series in $\alpha_s$ and $\Lambda_{\rm QCD}/m_b$
\cite{history}.
The leading order result in $1/m_b$ reproduces the parton model, while to
$\CO(1/m_b^2)$ two unknown nonperturbative parameters, $\bar\Lambda$ and
$\lambda_1$, arise.  The ratio $R_\tau$ defined in Eq.\ (\ref{rtau}) provides a
potential
constraint on these parameters,
although the uncertainty in the measurement is currently too large for these
constraints to
be useful  \cite{FLNN,Koyrakh,LN,FLS95}.
As in the case with massless leptons \cite{LSW} the $\CO(\alpha_s^2\beta_0)$
corrections to this process are quite large; however, these corrections largely
cancel in the ratio $R_\tau$.

We write the semitauonic decay of a $b$-quark in terms of the quark pole masses
$m_b$ and $m_c$ as
\begin{eqnarray}
\Gamma(\bxctaunu)&=&{G_F^2 m_b^5\over 192\pi^3}
\Gamma^{(0)}\left(\hat m_c,\hat m_\tau\right)\left(1+{\alpha_s(m_b)\over
\pi}\Gamma^{(1)}\left(\hat m_c,\hat m_\tau\right)
\vphantom{\left({\alpha_s(m_b)\over \pi}\right)^2}\right.\\
&&\qquad\left.
+\left({\alpha_s(m_b)\over \pi}\right)^2 \beta_0
\Gamma^{(2)}_\beta\left(\hat m_c,\hat m_\tau\right)+\dots\right) \nonumber
\end{eqnarray}
where $\hat m_c\equiv m_c/m_b$, $\hat m_\tau\equiv m_\tau/m_b$,
$\beta_0 = 11-{2\over 3}n_f$ is the QCD $\beta$-function, and
$n_f$ is the number of light quark flavors running through the vacuum
polarization loops.
The ellipsis denote terms ${\cal O}(\alpha_s^2)$ and higher.
The one-loop correction $\Gamma^{(1)}(\hat m_c,\hat m_\tau)$ is given in Ref.\
\cite{cjk}.

To compute the $\CO(\alpha_s^2 \beta_0)$ term $\Gamma^{(2)}_\beta$ we follow
the work of
Smith and Voloshin \cite{SmVol} and compute the
${\cal O}\left(\alpha_s\right)$ rate with a finite
gluon mass, $\Gamma^{(1)}(\hat m_g,\hat m_c,\hat m_\tau)$.
The ${\cal O}(\alpha_s^2 \beta_0)$ correction in the true theory,
$\Gamma_{\beta}^{(2)}(\hat m_c,\hat m_\tau)$,
can be found
from this rate by performing the weighted integral
\begin{eqnarray}\label{smitvol}
\Gamma_\beta^{(2)} (\hat m_c,\hat m_\tau)= -\beta_0{\alpha_s^{(V)} (m_b)\over
4\pi}
\int_0^\infty\ {dm_g^2\over m_g^2}
\left(  \Gamma^{(1)}(\hat m_g,\hat m_c,\hat m_\tau) - {m_b^2\over m_g^2+ m_b^2}
\Gamma^{(1)} (\hat m_c,\hat m_\tau)
\right)
\end{eqnarray}
where $\alpha_s^{(V)} (m_b)$ is the strong coupling defined in the
$V$-scheme of Ref.\ \cite{BLMa}, and is related to the coupling
$\alpha_s$ defined in the $\overline{MS}$ scheme by
\begin{equation}
\alpha_s^{(V)} (\mu) = \alpha_s(\mu)
+ {5\over 3}  { \alpha_s^2(\mu)\over 4\pi} \beta_0
+\dots\  .
\end{equation}
We have obtained a lengthy analytic expression for $d\Gamma(m_g)/dq^2$, where
$\sqrt{q^2}$ is the invariant mass of the lepton pair, which we have
integrated
numerically over $q^2$ and $m_g$ to obtain $\Gamma^{(2)}_\beta$.  Since the
results
are very sensitive to $\hat m_c$ and $\hat m_\tau$, we have chosen to follow
the
approach of Refs.\ \cite{FLS95} and express these
ratios as a power series in $1/m_{B}$:
\begin{eqnarray}\label{massratios}
  \hat m_c&=&{\bar m_D\over \bar m_B}-{\bar\Lambda\over m_B}\left(
  1-{\bar m_D\over \bar m_B}\right)-{\bar\Lambda^2\over m_B^2}\left(
  1-{\bar m_D\over \bar m_B}\right)+{\lambda_1\over 2 m_B m_D}\left(
  1-{\bar m_D^2\over \bar m_B^2}+\ldots\right) \\
  &=&0.372-0.628{\bar\Lambda\over m_B}-0.628{\bar\Lambda^2\over
  m_B^2}+1.16{\lambda_1\over m_B^2}\nonumber\\
  \hat m_\tau&=&{m_\tau\over \bar m_B}\left(1+{\bar\Lambda\over m_B}
  +{\bar\Lambda^2\over m_B^2}-{\lambda_1\over 2 m_B^2}+\ldots\right)\nonumber
\\
  &=&0.334+0.334{\bar\Lambda\over m_B}+0.334{\bar\Lambda^2\over
  m_B^2}-0.167{\lambda_1\over m_B^2}\,, \nonumber
\end{eqnarray}
where we have defined the spin-averaged meson masses
\begin{eqnarray}\label{massratio}
  \bar m_D\equiv{m_D+3 m_{D^*}\over 4}&=&m_c+\bar\Lambda-{\lambda_1\over
  2m_D}+\ldots\simeq 1975\,{\rm MeV}\, \\
  \bar m_B\equiv{m_B+3 m_{B^*}\over 4}&=&m_b+\bar\Lambda-{\lambda_1\over
  2m_B}+\ldots\simeq 5313\,{\rm MeV}.\nonumber
\end{eqnarray}
To the order in which we are working we can just use the
leading term in our perturbative calculation.  We find
\begin{eqnarray}\label{semictau}
\Gamma (B\to X_c\tau\overline{\nu}_\tau)
&=& |V_{bc}|^2\ {G_F^2 m_B^5\over 192\pi^3}\
[0.082]\left[1
 -1.94 {\overline{\Lambda}\over m_B}
 -  1.29 \left({\alpha_s(m_b) \over\pi}\right)
\right. \nonumber\\
& & \left.
- 1.28 \left({\alpha_s(m_b) \over\pi}\right)^2\beta_0
 +  {\cal O}\left( 1/m_B^2 , \alpha_s/m_B,\alpha_s^2\right)\right].
\end{eqnarray}
For completeness, we also give the result for $\hat m_c=0.3$ and
$m_b=4.80\,\mbox {GeV}$,
\begin{eqnarray}\label{semictau2}
\Gamma (B\to X_c\tau\overline{\nu}_\tau)
= |V_{bc}|^2\ {G_F^2 m_b^5\over 192\pi^3}\
[0.114]&&\left[1
 -  1.39 \left({\alpha_s(m_b) \over\pi}\right)
-1.58\left({\alpha_s(m_b) \over\pi}\right)^2 \beta_0\right.\nonumber\\
&&\left.\vphantom{\left({\alpha_s(m_b) \over\pi}\right)^2}\qquad+
{\cal O}\left( 1/m_b^2, \alpha_s^2\right)\right].
\end{eqnarray}

As is the case for $b\rightarrow ce\bar \nu_e$ decays, the
$\CO(\alpha_s^2\beta_0)$
corrections in Eqs.~(\ref{semictau}) and (\ref{semictau2}) are
quite large, corresponding to a low BLM scale for this process.  However, these
corrections largely drop out of the ratio $R_\tau$.  Combining
Eq.~(\ref{semictau})
with the results  of \cite{LSW}, we find
\begin{eqnarray}\label{rtaub}
R_\tau  &=&  0.224\left[
1  +  0.24 {\alpha_s(m_b)\over\pi}
 +  0.15 \left({\alpha_s(m_b)\over\pi}\right)^2\beta_0
- 0.29 {\overline{\Lambda}\over m_B}
 +  {\cal O}\left( 1/m_B^2 , \alpha_s^2\right)
\right]\\
&\simeq &0.224\left[ 1+0.017+0.007- 0.29 {\overline{\Lambda}\over m_B}
 +  {\cal O}\left( 1/m_B^2 , \alpha_s^2\right)
\right]\nonumber
\end{eqnarray}
where we have taken $\alpha_s(m_b)=0.23$ in the second line.
The perturbation series appears well behaved, and
the corresponding BLM scale for $R_\tau$ is $\mu_{\rm BLM}= 0.29\, m_b$.

\bigskip
\section{${\cal O}(\alpha_s^2\beta_0)$ corrections to $\bccs$}
\medskip

Neglecting the $s$ quark mass\footnote{Since $m_s\sim \Lambda_{\rm QCD}$, we
will treat terms of order $m_s^2$ to be of the same size as terms of order
$\Lambda_{\rm QCD}^2$, which we are neglecting.}, we write the width for
$\bccs$ decays
(where the final state includes an arbitrary number of gluons and light quarks)
as
\begin{equation}
\Gamma(\bccs)={G_F^2 m_b^5\over 64\pi^3}
\Gamma^{(0)}\left(\hat m_c\right)\left(1+{\alpha_s(m_b)\over
\pi}\Gamma^{(1)}\left(\hat m_c\right)+
\left({\alpha_s(m_b)\over \pi}\right)^2
\Gamma^{(2)}\left(\hat m_c\right)+\dots\right)
\end{equation}
where
\begin{equation}
\Gamma^{(0)}(x)=\sqrt{1-4 x^2}\left(1-14 x^2-2 x^4-12 x^6\right)+24 x^4
\left(1-x^4\right)\ln\left({1+\sqrt{1-4 x^2}\over 1-\sqrt{1-4 x^2}}\right)
\end{equation}
is the tree-level result; $\Gamma^{(0)}(0.30)=0.196$.  The complete one-loop
corrections may be obtained from
Refs.\ \cite{mv,bbfg}; for $\hat m_c=0.30$ one obtains
$\Gamma^{(1)}(0.3)=2.99$.  Taking $\alpha_s(m_b)=0.23$, this corresponds to a
$22\%$ enhancement of the rate over the tree level result.

Since the four-quark operators responsible for nonleptonic $b$ decays run in
the effective
theory below $m_W$ the $\CO(\alpha_s^2)$ contributions to the decay are more
complicated than for semileptonic decays.  We write the
$\CO(\alpha_s^2)$ contribution to the decay rate as
\begin{equation}
\Gamma^{(2)}(\hat m_c)=c_1 \ln^2{m_W\over m_b}+c_2(\hat m_c)\ln{m_W\over m_b}+
c_3(\hat m_c)\beta_0+c_4(\hat m_c)
\end{equation}
where $c_1=4$ \cite{alta}.
The subleading log contribution $c_2(\hat m_c)$ was calculated in Ref.\
\cite{bbfg}; for
$\hat m_c=0.30$ these authors find $c_2=3.34$.

Clearly the requirement that $c_3$ dominates the two-loop
correction, implicit in the BLM approach,
will not hold in this process, since the non-vacuum polarization
terms
$c_1$ and $c_2$ are enhanced by powers of $\ln m_W/m_b$.  Separating these
terms out, we may instead hope that $c_3$ dominates over $c_4$ due to the
factor of $\beta_0$.  However, even this assumption may not hold.
Voloshin \cite{mv} has shown that in the limit
in which the charm is produced nearly at rest $c_4$
receives a large enhancement.  For $\bcud$, in this limit the analog of $c_4$
is of
order $\pi^2$,  whereas for $b\rightarrow c\bar c s$ the Coloumb
exchange graphs between the two slowly-moving charmed quarks give a
contribution to $c_4$ of order $\pi^4$.  While it is not known whether this
enhancement is relevant for the physical value of
$\hat m_c$, it indicates that the
$\CO(\alpha_s^2\beta_0)$ terms need not dominate over the $\CO(\alpha_s^2)$
terms.
Nevertheless, as a first step towards understanding the size of the two-loop
corrections to
this process, we may calculate $c_3$.

While the complete series of leading and subleading logs has
been summed to all orders \cite{alta,bbfg}, we cannot consistently use these
results
since we are not summing all terms of
$\CO(\alpha_s^n\log^{n-2}(m_b/m_W)\beta_0)$.
However, as was stressed in Ref.\ \cite{mv2}, $\ln(m_W/m_b)\approx 2.8$ is not
a large number, and the leading log expansion does not seem to work well for
nonleptonic
$b$ decays.  For example, for $b\rightarrow c\bar u d$ decay the subleading
$\CO\left(\alpha_s^2\ln(m_W/m_b)\right)$ term is 2/3 the size of the leading
$\CO\left(\alpha_s^2\ln^2(m_W/m_b)\right)$ term.
Therefore, we choose to work consistently to
$\CO(\alpha_s^2)$ and discard the rest of the leading and subleading log terms.
The neglected terms of $\CO(\alpha_s^3)$ and
above are likely
to be much smaller than the uncomputed $\CO(\alpha_s^2)$ corrections.

The calculation of $c_3$ is simplified due to the fact that the graphs
factorize into the contribution from the upper $b c$ vertex
and the contribution from the lower $\bar cs$ vertex.
The upper vertex contribution can be simply obtained from the
corrections to
$\bctaunu$\ (by making the substitution $m_\tau \to m_c$), while the
contribution from the lower vertex require an additional calculation.

For the lower vertex corrections, the kinematic structure of the phase space
allows us to express the integrals over
the momenta of the $\bar{c}$ quark, $s$ quark, and
gluon in terms of the spectral density of the charged
V-A current (the imaginary part of the charged current vacuum
polarization),
\begin{equation}
\Gamma^{(1)}_{\rm lower}(m_g) = \int{{\cal M}_\mu {\cal M}^{*}_\nu \
 {\rm Im\,}{\Pi_{\mu\nu}(q^2)} \ {(2 \pi)^3 \over{2 m_b}} \sqrt{s} \ d{\tau}_2
(P_b;
p_c,q) \ dq^2} ,
\label{fullcorrection}
\end{equation}
where ${\cal M}_\mu$ is the contribution from the $bc$ line and
${\rm Im\,}{\Pi_{\mu\nu}}(q^2)$ is the imaginary part of the vacuum
polarization. The tensor structure of the vacuum polarization can be
decomposed into a transverse and a longitudinal contribution,
\begin{equation}
{\rm Im\,}{\Pi_{\mu\nu}(q^2)} = {q_{\mu} q_{\nu}\over{q^2}} P_{l}(q^2)+ \left
  ( {g_{\mu\nu}}-{q_{\mu} q_{\nu}\over{q^2}} \right ) P_{t}(q^2) .
\end{equation}
Since the functions $P_{l}(q^2)$ and $P_{t}(q^2)$ depend only on
the scalar $q^2$, the integration over $d{\tau}_2 (P_b;p_c,q)$ can
be carried out analytically with a simple computation,
\begin{eqnarray}
\Gamma^{(1)}_{\rm lower}(m_g) = { 16 \pi \over{m_b^7}} \int{ \left \{
\left [ (m_b^2-m_c^2)^2-q^2
    (m_b^2+m_c^2) \right ] P_l (q^2) + \right. }\nonumber \\
\left.\left [ (m_b^2+m_c^2)^2+ q^2 (m_b^2+m_c^2-2 q^2)\right ] P_t (q^2)
\right \}
\sqrt{\lambda(m_b^2,m_c^2,q^2)}{dq^2 \over{\sqrt{q^2}}}
\label{finalcorrection}
\end{eqnarray}
where $\lambda(x,y,z)=x^2+y^2+z^2-2 xy-2 xz-2 yz$.  The resulting expression
is quite lengthy and we do not present it here.
The functions $P_{t} (q^2)$ and $P_{l}(q^2)$ have been previously calculated
(for a massless gluon) to ${\cal
O}(\alpha_s)$ in the context of QCD sum rules
\cite{reindersetal}.

\begin{figure}
\epsfxsize=16cm
\hfil\epsfbox{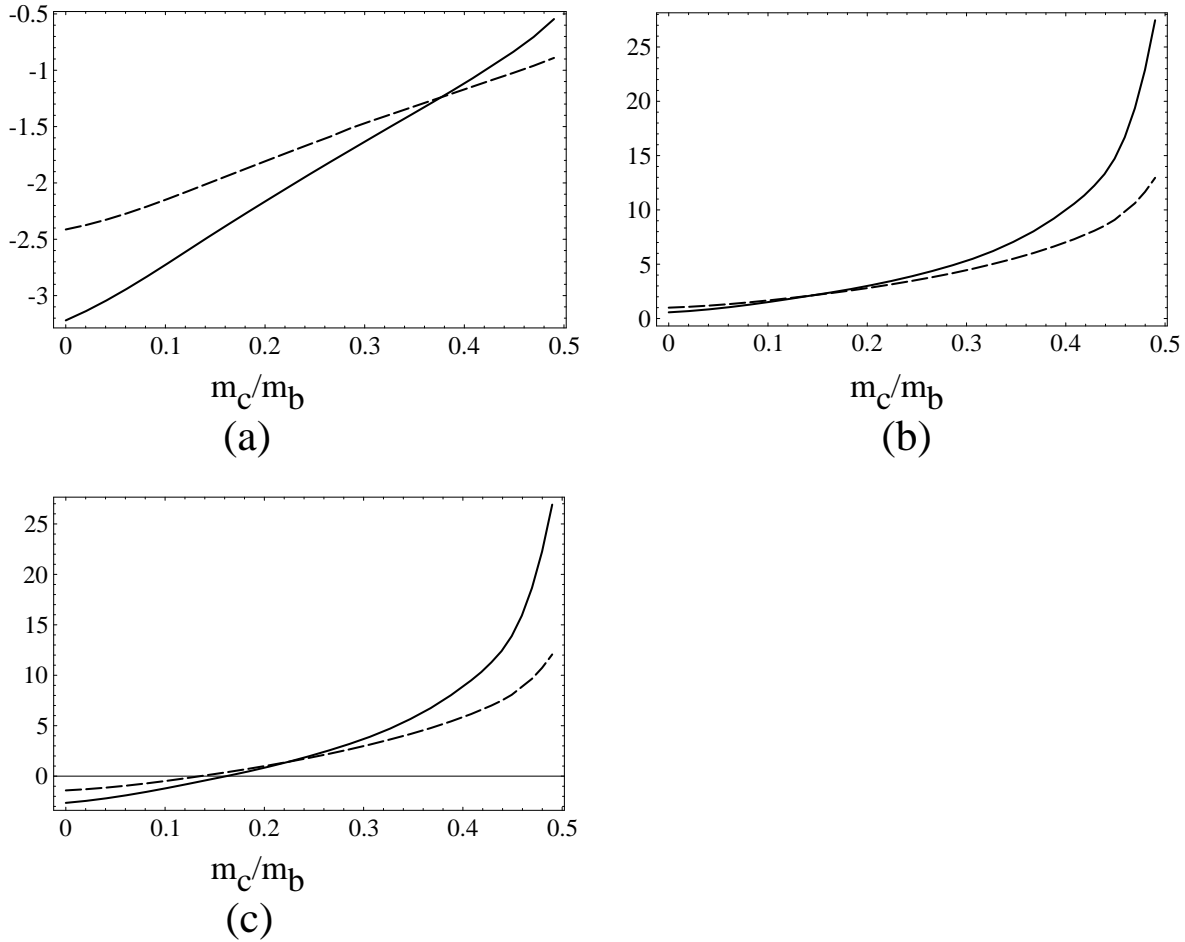}\hfill
\caption{Contributions to $\Gamma^{(1)}$ (dashed lines) and $c_3$
(solid lines) from (a) the renormalization of
the $b c$ vertex, (b) the renormalization of the $\bar c s$ vertex, and
(c) the sum of (a) and (b).}
\label{scalefigs}
\end{figure}

\begin{figure}
\epsfxsize=12cm
\hfil\epsfbox{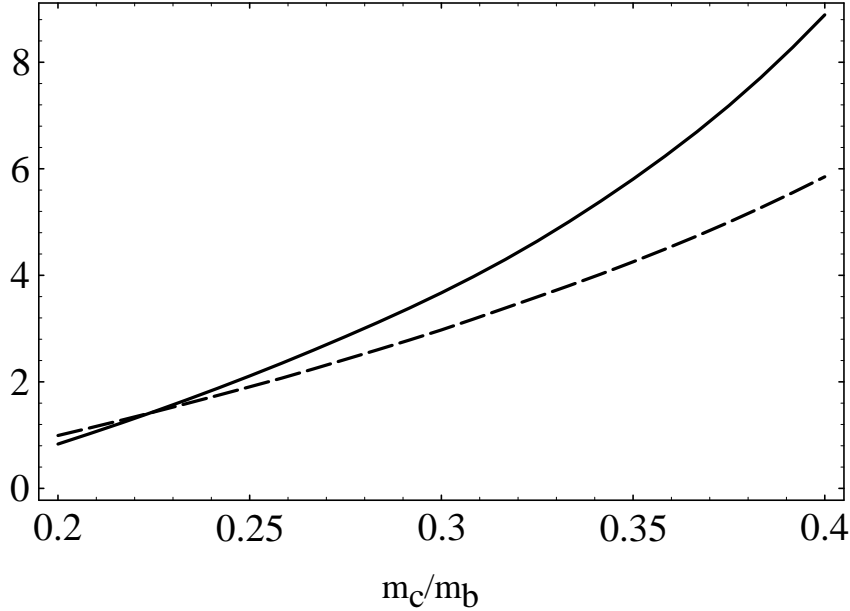}\hfill
\caption{$\Gamma^{(1)}$ (dashed line) and $c_3$
(solid line) as functions of $\hat m_c$ (expanded view of
Figure 1 (c)).}
\label{scalefigs2}
\end{figure}

It is then a simple matter to integrate numerically the resulting expression
over $q^2$
and $m_g^2$ to obtain $c_3$ as a function of $\hat m_c$.
At the ``reference point" $\hat m_c=0.3$, we find $c_3(0.3)=3.7$.
Using $\alpha_s(m_b)=0.23$ and $\beta_0=9$, this corresponds to an 18\%
correction to the tree level result, almost as large as the one loop
correction.
The values of $c_3(\hat m_c)$ for a range of values of $\hat m_c$ are given
in Table
\ref{scaletable} and plotted in Figs. \ref{scalefigs} and \ref{scalefigs2}
along with the separate contributions
from the upper and lower vertices.
\begin{table}
$$
\begin{array}{c|cccc}
\hat m_c &\ \ \Gamma^{(1)}(\hat m_c)\ \  &\ \  c_3(\hat m_c)\ \  &
\ \ {\alpha_s(m_b)\over \pi}\Gamma^{(1)}(\hat m_c)\ \
&\ \ \left({\alpha_s(m_b)\over\pi}\right)^2\beta_0 c_3(\hat m_c)\ \ \\ \hline
0.20 & 0.99 & 0.83 & 0.07 & 0.04 \\
0.25 & 1.91  & 2.11& 0.14 & 0.10 \\
0.30 & 2.97  & 3.67& 0.22 & 0.18 \\
0.35 & 4.25 & 5.81& 0.31 & 0.28\\
0.40 & 5.85 & 8.89& 0.43 & 0.43
\end{array}
$$
\caption{Numerical values of the one and partial two loop corrections
$\Gamma^{(1)}$and $c_3$
for $\bccs$ decay.  In the last two columns we have
taken $\alpha_s(m_b)=0.23$ and $\beta_0=9$.}
\label{scaletable}
\end{table}

It is useful to compare these results with the leading and subleading log
corrections to $\Gamma^{(2)}$.  For $\hat m_c=0.3$, these are
\begin{equation}
\left(4\ln^2 {m_W\over m_b}+3.34\ln {m_W\over m_b}\right)
\left({\alpha_s\over\pi}\right)^2\simeq (3.5+1.0)\beta_0
\left({\alpha_s\over\pi}\right)^2
\end{equation}
where we have removed a factor of $\beta_0=9$ to allow comparison with the
second
column in Table \ref{scaletable}.
For $\hat m_c=0.3$ the $\alpha_s^2\beta_0$ term is
roughly the same size as the $\CO(\alpha_s^2)$ leading log correction, and a
factor
of three greater than the $\CO(\alpha_s^2)$ subleading log.

Note that the $\CO(\alpha_s)$ and $\CO(\alpha_s^2\beta_0)$ corrections to the
$\bar cs$
vertex are  positive for all values of $\hat m_c$,
while the corrections to the $bc$ vertex are negative.  The one loop
corrections
cancel at $\hat m_c\approx 0.14$, while the $\CO(\alpha_s^2\beta_0)$
corrections
cancel at a slightly higher value of $\hat m_c$.
In this situation
the BLM scale $\mu_{\rm BLM}$ is
not physically relevant: at the point where the one loop corrections to the
vertices cancel, $\mu_{\rm BLM}$ is singular, whereas at the point where the
$\CO(\alpha_s^2\beta_0)$ contributions cancel $\mu_{\rm BLM}=m_b$.
In this region the BLM
scale for the decay width is unrelated to the BLM scales that would be
obtained for the upper and lower
vertices individually, and does not reflect the average momentum of the gluons
in
the diagrams.  Therefore we prefer simply to present our results as a
contribution to the
$\CO(\alpha_s^2)$ correction to the decay rate.

We also note that since the leading order phase space function
$\Gamma^{(0)}(\hat m_c)$ is very
sensitive to the $b$ and $c$ quark masses, there is a large uncertainty in the
total
$\bccs$ width simply due to the uncertainty in the $b$ and $c$ quark masses.
Since $m_b$ and $m_c$ are related via the $1/m_Q$ expansion
to the corresponding hadron masses, this sensitivity is really an additional
hidden source of
$1/m_Q$ corrections, just as in the semileptonic decay width.
This is made clear if we adopt the approach of the previous section
and write $\hat m_c$ as a series in $1/m_{B}$.  In this case, the large
sensitivity to
$\hat m_c$ results in a $1/m_B$ correction which is as large as the leading
order term,
\begin{equation}\label{hadbccs}
\Gamma (B\to X_{c\bar cs})
= |V_{bc} V_{cs}|^2\ {G_F^2 m_B^5\over 64\pi^3}
 \left[  0.057 \right]
\left[1 + 9.7 {\overline{\Lambda}\over m_B}
 +   {\cal O}\left( 1/m_B^2 , \alpha_s\right)\right].
\end{equation}
Of course, one could argue that this result is misleading because we are
expanding about
the extreme value $\hat m_c=0.37$.   Nevertheless, the large $1/m_b$ correction
shows the sensitivity of the width to the quark masses.  Working instead with
pole
masses and keeping $m_b$ fixed,
varying $\hat m_c$ between 0.27 and 0.32 results in a factor of two change in
the total rate.

It is straightforward to find the $\alpha_s^2\beta_0$ term for the decay
$b\rightarrow c\overline{u}d$ from computations of the charmed semileptonic
decay
\cite{LSW} and from the results for $R_{e^+e^-}$ \cite{vacpol}.  For $\hat
m_c=0.3$
this gives
\begin{eqnarray}\label{hadbcud}
\Gamma (\bcud)
&=& |V_{bc} V_{ud}|^2\ {G_F^2 m_b^5\over 64\pi^3}\
[0.52] \left[1
- 0.67 {\alpha_s (m_b) \over\pi}
+4\ln^2 {m_W\over m_b}\left({\alpha_s (m_b) \over\pi}\right)^2\right.
\nonumber\\
&&\qquad \left.
+7.17\ln {m_W\over m_b}\left({\alpha_s (m_b) \over\pi}\right)^2- 1.1
1\left({\alpha_s (m_b) \over\pi}\right)^2\beta_0
+  {\cal O}\left(\alpha_s^2\right)
\right].
\end{eqnarray}
Combining this with the results of the present work, we find the ratio of
the partial widths for $\hat m_c=0.3$,
\begin{eqnarray}
{\Gamma(\bccs)\over\Gamma(\bcud)}&=&0.376{|V_{cs}|^2\over |V_{ud}|^2}
\left[1+3.66{\alpha_s (m_b) \over\pi}+ 4.80\beta_0\left({\alpha_s (m_b)
\over\pi}\right)^2
\right. \\
&&\left.\qquad-3.83\ln {m_W\over m_b}\left({\alpha_s (m_b)
\over\pi}\right)^2+\dots\right]\nonumber
\end{eqnarray}
The $\alpha_s^2\beta_0$ correction enhances the tree-level ratio by  $22\%$.

Similarly, the $\CO(\alpha_s^2\beta_0)$ enhancement of $\Gamma(\bccs)$ will
decrease the semileptonic branching fraction and increase the charm
multiplicity
$\langle n_c\rangle$.    Combining the result for $\bccs$ with the
$\CO(\alpha_s^2\beta_0)$
corrections to the other modes, we find for $\hat m_c=0.3$, an
$\CO(\alpha_s^2\beta_0)$
correction shift to the semileptonic branching fraction of
\begin{equation}\label{semibrth}
\delta\left({\Gamma_{\rm s.l.}\over \Gamma}\right)=
-0.19\, \beta_0\left({\alpha_s (m_b) \over\pi}\right)^2=-0.009.
\end{equation}
The corresponding shift to the charm multiplicity $\langle n_c \rangle$ is
\begin{eqnarray}
\delta \langle n_c\rangle&=&0.74 \beta_0\left({\alpha_s (m_b)
\over\pi}\right)^2=0.036.
\end{eqnarray}
Since
we are simply illustrating the effect of the $\CO(\alpha_s^2\beta_0)$ terms on
these
observables, we do not include the remaining perturbative corrections or 
contributions from rare decay modes in these
expressions.

\bigskip
\section{Conclusions}
\medskip

We have computed the ${\cal O}\left(\alpha_s^2\beta_0\right)$ contributions
to the rate of the nonleptonic decay $\bccs$ at the parton level.   While these
corrections
do not dominate in any formal limit of the theory, they are a well-defined
subset of the
complete two-loop corrections.   When the perturbation series is expressed in
terms
of  $\alpha_s(m_b)$, the $\CO(\alpha_s^2\beta_0)$ corrections are of the same
order as
both the one-loop corrections and the leading log corrections.
For $\hat m_c=0.3$ they provide
an
additional reduction of $\sim 1\%$ in the semileptonic branching fraction,
and
increase the charm multiplicity $\langle n_c\rangle$ by $\sim 0.04$.

These corrections are sufficiently large to cast doubt on the applicability of
perturbative QCD to this decay mode.  Since there is so little phase space,
this is
not unexpected.  These corrections are in addition to the large
$\CO(\alpha_s^2)$
corrections suggested by Voloshin \cite{mv}, as well as the large implicit
$1/m_{b,c}$
corrections due to the uncertainties in the $c$ and $b$ masses.

\acknowledgements

This work was
supported by the United States Department of Energy under Grant
No.~DE-FG02-91ER40682 and by the Natural Sciences and Engineering
Research Council of Canada.
M.J.S. acknowledges additional
support from the United States Department of Energy for Outstanding
Junior Investigator Award No.~DE-FG02-91ER40682.

\end{document}